\documentclass[journal,a4paper,onecolumn]{IEEEtran}

\usepackage{tikz}
\usepackage{epsfig}
\usepackage{epstopdf}
\usepackage{siunitx}
\usepackage{pbox}
\usepackage{makecell}
\usepackage{multirow}
\usepackage{amssymb}
\usepackage{amsmath}
\usepackage[normalem]{ulem}

\usepackage{blindtext}
\usepackage{etoolbox}
 
\makeatletter

\docsvlist{\@oddhead,\@evenhead,\ps@headings,\ps@IEEEtitlepagestyle,\ps@IEEEpeerreviewcoverpagestyle}
\makeatother

\begin{document}

\title{An extraction method for mobility degradation and contact resistance of graphene transistors}

\author{Anibal Pacheco-Sanchez, Nikolaos Mavredakis, Pedro C. Feijoo, David Jiménez 

\thanks{This work has received funding from the European Union’s Horizon 2020 research and innovation programme under grant agreements No GrapheneCore3 881603, from Ministerio de Ciencia, Innovación y Universidades under grant agreement RTI2018-097876-B-C21(MCIU/AEI/FEDER, UE) and FJC2020-046213-I. This article has been partially funded by the European Union Regional Development Fund within the framework of the ERDF Operational Program of Catalonia 2014-2020 with the support of the Department de Recerca i Universitat, with a grant of 50\% of total cost eligible. GraphCAT project reference: 001-P-001702.\newline \indent The authors are with the Departament d'Enginyeria Electr\`{o}nica, Escola d'Enginyeria, Universitat Aut\`{o}noma de Barcelona, Bellaterra 08193, Spain, e-mails: AnibalUriel.Pacheco@uab.cat, David.Jimenez@uab.cat}
}
\maketitle
\makeatletter
\def\ps@IEEEtitlepagestyle{
  \def\@oddfoot{\mycopyrightnotice}
  \def\@evenfoot{}
}
\def\mycopyrightnotice{
  {\footnotesize
  \begin{minipage}{\textwidth}
  \centering
© 2022 IEEE.  Personal use of this material is permitted.  Permission from IEEE must be obtained for all other uses, in any current or future media, including reprinting/republishing this material for advertising or promotional purposes, creating new collective works, for resale or redistribution to servers or lists, or reuse of any copyrighted component of this work in other works. DOI: 10.1109/TED.2022.3176830
  \end{minipage}
  }
}

\begin{abstract}
\boldmath
The intrinsic mobility degradation coefficient, contact resistance and the transconductance parameter of graphene field-effect transistors (GFETs) are extracted for different technologies by considering a novel transport model embracing mobility degradation effects within the charge channel control description. By considering the mobility degradation-based model, a straightforward extraction methodology, not provided before, is enabled by applying the concept of the well-known \textit{Y}-function to the \textit{I-V} device characteristics. The method works regardless the gate device architecture. An accurate description of experimental data of fabricated devices is achieved with the underlying transport equation by using the extracted parameters. An evaluation of the channel resistance, enabled by the extracted parameters here, has been also provided.  

\end{abstract}
%
%

\IEEEpeerreviewmaketitle
\section{Introduction}
\label{ch:intro}

One of the most popular approaches for the characterization of graphene field-effect transistors (GFETs) has been the description of experimental data by a device total resistance $R_{\rm tot}$ model \cite{KimNah09}. This model considers the impact of vertical fields and impurities on the carrier transport and has enabled to obtain relevant device parameters since its proposal, e.g., a low-field mobility \cite{LiaBai10}-\cite{MavWei20}. Recently, this well-known model has been adapted in order to explicitly consider mobility degradation effects in the performance description of GFET technologies \cite{JepAsa21}. This novel approach has been motivated by experimental observations of a carrier density-dependent mobility in GFETs under different scenarios \cite{ZhuPer09}-\cite{ZhoZha15}. Hence, a first-order mobility model considers both, short- and long-range scattering mechanisms in graphene devices by including an intrinsic mobility degradation coefficient $\theta_{\rm ch}$ into the mobility description, similarly to the approach used in MOSFET theory \cite{MerBor72}-\cite{EnzVit06}. Additionally, a potential drop at the metal-graphene interface has been considered by a contact resistance $R_{\rm c}$ in order to describe the internal transport phenomena. 

From a device characterization point of view, the fitting procedure relying on an initial guess of $R_{\rm c}$ towards obtaining the other three model parameters in \cite{JepAsa21} might become an important constraint for the characterization of GFETs at a large scale due to the computational burden and challenges related to randomly finding adequate initial values. A straightforward parameter extraction methodology considering experimental data is required to expand and ease the use of this novel modeling-based characterization approach. In this work, by following a methodical treatment of experimental data based on the concept of the well-known $Y$-function, a parameter extraction methodology for the mobility degradation-based transport model is proposed using individual device transfer characteristics of a given GFET technology with devices of different gated-channel lengths. In contrast to the original approach \cite{JepAsa21}, the extraction methodology proposed here is more robust for device characterization purposes since it yields values of three out of four of the model parameters without needing any initial random assumption. The fourth parameter, related to charge-neutrality, is extracted by appropriately adjusting the model to experiments at very low transversal electric fields.

By considering the characterization of GFET technologies including sub-\SI{}{\micro\meter} gate-lengths, the model-based characterization is proven here to be for general purposes and not only for long devices as shown in \cite{JepAsa21}. Further adaptations of the underlying model to include velocity saturation effects -relevant at high lateral fields- are out of the scope of this work, however, the extracted parameters here can be relevant for the extraction of velocity saturation related parameters. In contrast to previous GFET studies using \textit{Y}-function-based methods \cite{UrbLup20}-\cite{PacJim21}, this work relies on a transport current description including a charge definition more related to GFETs physics rather than the conventional MOSFET charge equation. The extracted parameters enable an evaluation of the mobility and an accurate description of the drain current over a wide transversal electric field range including charge neutrality conditions. 

This straightforward extraction methodology of contact resistance and mobility parameters of graphene technologies can be applied to any gate architecture. Hence, it is an immediate efficient alternative to test-structure-based extraction methodologies such as the transfer length method (TLM) \cite{Sch06}, limited to global-back-gate (GBG) devices only. The latter limitation can mislead performance projections based on TLM for other device architectures in which the different electrostatics, e.g., from a top- (TG) \cite{HanOid13} or buried-gate (BG) \cite{LeeKim19}, might affect the Schottky-like potential barrier height at graphene-metal contacts \cite{ChaJim15}, \cite{WanMal21}, i.e., the value of $R_{\rm c}$. Furthermore, the fabrication issues of individual TG and BG \cite{LeeKim19} graphene devices, implies a more challenging fabrication of TLM structures (including a series of individual devices) for such configurations. The TLM-related constraints are overcome by the method presented here since (i) the particular device electrostatics is considered in the underlying transport description and (ii) no fabrication of extra test structures is required.

\section{Parameters extraction methodology}


The drain current $I_{\rm D}$ of GFETs given by a straightforward adaptation of a widely used \textit{I-V} model \cite{KimNah09} considering both an appropiate charge control description and an explicit contribution of the extrinsic mobility degradation coefficient $\theta$ \cite{Ris83}, \cite{HaoCab85}, \cite{Ghi88} is expressed as\footnote{Notice that $\theta$ has been called $\theta_{\rm eff}$ in \cite{JepAsa21}.} \cite{JepAsa21}

\begin{equation}
I_{\rm D} = \beta \frac{\sqrt{V_0^2 + V_{\rm GSO}^2}}{1 + \theta \sqrt{V_0^2 + V_{\rm GSO}^2}} V_{\rm DS},
\label{eq:Id}
\end{equation}

\noindent where $\beta$ is a transconductance parameter associated to $\mu_{\rm app,0} C_{\rm{ox}}w_{\rm{g}}/L_{\rm g}$ \cite{Ghi88} with $\mu_{\rm app,0}$ as an apparent low-field mobility, $C_{\rm{ox}}$ as the gate oxide capacitance per unit area, $w_{\rm{g}}$ as the gate width and $L_{\rm g}$ as the gated-channel length, $V_{\rm GSO}=V_{\rm GS}-V_{\rm Dirac}$ is the extrinsic gate-to-source voltage overdrive with $V_{\rm GS}$ as the extrinsic gate-to-source voltage and $V_{\rm Dirac} (= V_{\rm GS}\vert_{\min (I_{\rm D})})$ as the $V_{\rm GS}$ corresponding to charge neutrality conditions, $V_0 = qn_0/C_{\rm ox}$ the residual voltage with $q$ as the electron charge, $n_0$ as the residual charge carrier density at $V_{\rm Dirac}$ \cite{KimNah09}, $\theta = \theta_{\rm ch} + \beta R_{\rm c}$ with $\theta_{\rm ch}$ as the intrinsic mobility degradation coefficient due to vertical fields \cite{MerBor72}-\cite{EnzVit06} and the contact resistance $R_{\rm c}(=R_{\rm cs}+ R_{\rm cd})$ embracing the phenomena at metal-channel interfaces (source and drain sides). Notice that Eq. (\ref{eq:Id}) can be also obtained from the drift-diffusion transport description for graphene transistors \cite{MavWei19} after neglecting the quantum capacitance and the local channel voltage at low $V_{\rm DS}$.

Based on Eq. (\ref{eq:Id}) and by considering that $V_{\rm GSO} \gg V_{\rm 0}$, the transconductance $g_{\rm m}=\partial I_{\rm D}/ \partial V_{\rm GS}$ is given by

\begin{equation}
g_{\rm m} \approx \beta \frac{1}{\left(1+\theta \sqrt{V_0^2 + V_{\rm GSO}^2}\right)^2} V_{\rm DS},
\label{eq:gm}
\end{equation}

\noindent while the device total resistance $R_{\rm tot}=V_{\rm DS}/I_{\rm D}$, considering the definition of $\theta$, reads

\begin{equation}
R_{\rm tot} = \underbrace{\left(\frac{1}{\beta \sqrt{V_0^2 + V_{\rm GSO}^2}} + \frac{\theta_{\rm ch}}{\beta}\right)}_{R_{\rm ch}} + R_{\rm c}.  
\label{eq:Rtot}
\end{equation}

\noindent A comparison with the general definition of $R_{\rm tot}$ as the sum of channel resistance $R_{\rm ch}$ and $R_{\rm c}$, and by considering that $\theta_{\rm ch}$ is associated to channel phenomena \cite{MerBor72}-\cite{EnzVit06}, leads to relate the term inside brackets in Eq. (\ref{eq:Rtot}) as $R_{\rm ch}$ in the context of this work. The latter is confirmed by considering two channel scattering mechanisms (cf. Eq. (3) in \cite{JepAsa21}) in the underlying transport model (cf. Eq. (3) in \cite{KimNah09})\footnote{$R_{\rm tot}=\left(N_{\rm sq}/\mu_0\right)\left(1/qn + \theta_{\rm ch}/C_{\rm ox}\right) + R_{\rm c} \equiv R_{\rm ch} + R_{\rm c}$, with the number of squares $N_{\rm sq}=L_{\rm g}/w_{\rm g}$}.

By applying the concept of the $Y$-function \cite{Ghi88}, \cite{Jai88}, i.e., $Y=I_{\rm D}/\sqrt{g_{\rm m}}$, with Eqs. (\ref{eq:Id}) and (\ref{eq:gm}), this relation yields

\begin{equation}
Y \approx \sqrt{\beta V_{\rm DS}} \sqrt{V_0^2 + V_{\rm GSO}^2},
\label{eq:Yfun}
\end{equation}

\noindent while the product of Eqs. (\ref{eq:Rtot}) and (\ref{eq:Yfun}) corresponds to

\begin{equation}
R_{\rm{tot}} Y \approx \sqrt{\frac{V_{\rm{DS}}}{\beta}} + \frac{\theta}{\beta}Y.
\label{eq:RtotY}
\end{equation} 

\noindent The ratio $\theta/\beta$ and the parameter $\beta$ can be extracted from the plots of $R_{\rm tot}Y$ vs. $Y$ and $Y^2$ vs. $V_{\rm GSO}^2$, respectively, for different $L_{\rm g}$ and at a fixed low $V_{\rm DS}$. By following the definition of $\theta$ given above, $\theta_{\rm ch}$ and $R_{\rm c}$ are extracted from the intercept at zero and slope of a $\beta$-dependent plot of $\theta$. While the extraction of $\beta$, $\theta_{\rm ch}$ and $R_{\rm c} $ is performed at an unipolar bias region away from $V_{\rm Dirac}$, the underlying model can be adjusted towards describing charge neutrality conditions. Hence, a unique value for $V_0$ is  obtained from the best adjustment of the model to experimental data at the bias region close to $V_{\rm Dirac}$, upon extraction of all other parameters. 

The extraction method yields a constant $R_{\rm c}$ value which can be of special interest for technology evaluation \cite{WuTse15} as well as an initial guide for modeling purposes towards the design of specific applications at a fixed bias point, e.g., high-frequency circuits\footnote{In general, GFETs are fixed at $\vert V_{\rm GS} - V_{\rm Dirac}\vert \gg \SI{0}{}$ in practical scenarios where a high transconductance in an unipolar region benefits the dynamic performance.} \cite{PacRam21}, \cite{HamAsa21}. For the interested reader, a bias-dependent contact resistance model, which is out of the scope of this work, has been reported elsewhere \cite{ChaJim15}. Notice that despite Eq. (\ref{eq:RtotY}) resembles the approach presented in \cite{TreTre17} for silicon technologies, both the underlying transport equation and the extraction methodology are different in this work.

\section{Characterization of GFET technologies}

The parameter extraction methodology described above has been applied to the experimental data of GFET technologies with different device architectures such as GBG \cite{HanChe11}, \cite{GahKat20}, BG \cite{HanOid13}, \cite{MavWei19} and TG \cite{BaiLia11} as well as with different sub-\SI{}{\micro\meter}- and \SI{}{\micro\meter}-long gate lengths. Device characteristics and extracted parameters are listed in Table \ref{tab:results}. 

\begin{table} [!htb] 
\begin{center}
\caption{Characteristics and extracted model parameters of  GFETs.}
\begin{tabular}{c|c|c||c|c}

\makecell{gate arch.\\ \& ref.} & \makecell{$w_{\rm g}$\\$(\SI{}{\micro\meter})$} & \makecell{$L_{\rm g}$\\$(\SI{}{\micro\meter})$} & \makecell{$\theta_{\rm ch}$\\$(\SI{}{\volt^{-1}})$} & \makecell{$R_{\rm c}\cdot w_{\rm g}$\\ $(\SI{}{\kilo\ohm\cdot\micro\meter})$}   \\ \hline \hline

GBG, \cite{HanChe11} & \SI{5}{} &\makecell{\SI{0.120}{}, \SI{0.225}{},\\\SI{0.640}{}, \SI{2}{}} & \SI{0.025}{} & \SI{2.9}{}  \\

GBG, \cite{GahKat20} & \SI{20}{} &\makecell{\SI{5}{}, \SI{10}{}, \SI{15}{}\\\SI{20}{}, \SI{25}{}, \SI{30}{}} & \SI{0.034}{} & \SI{1.6}{}  \\

BG, \cite{HanOid13} & -- & \makecell{\SI{0.250}{},\\ \SI{0.450}{}} & \SI{1.5}{} & $\SI{0.2}{}\cdot w_{\rm g}$  \\

BG, \cite{MavWei19} & \SI{12}{} &\makecell{\SI{0.1}{}, \SI{0.2}{},\\\SI{0.3}{}} & \SI{3.1}{} & \SI{2.6}{}  \\

TG, \cite{BaiLia11} & \SI{3.4}{} &\makecell{\SI{0.1}{}, \SI{0.3}{},\\\SI{1}{}, \SI{2}{}} & \SI{0.081}{} & \SI{0.4}{}  \\

\end{tabular} \label{tab:results}
\end{center}
\end{table}  
  
\begin{figure}[!hbt]
	\centering
	\includegraphics[height=0.245\textwidth]{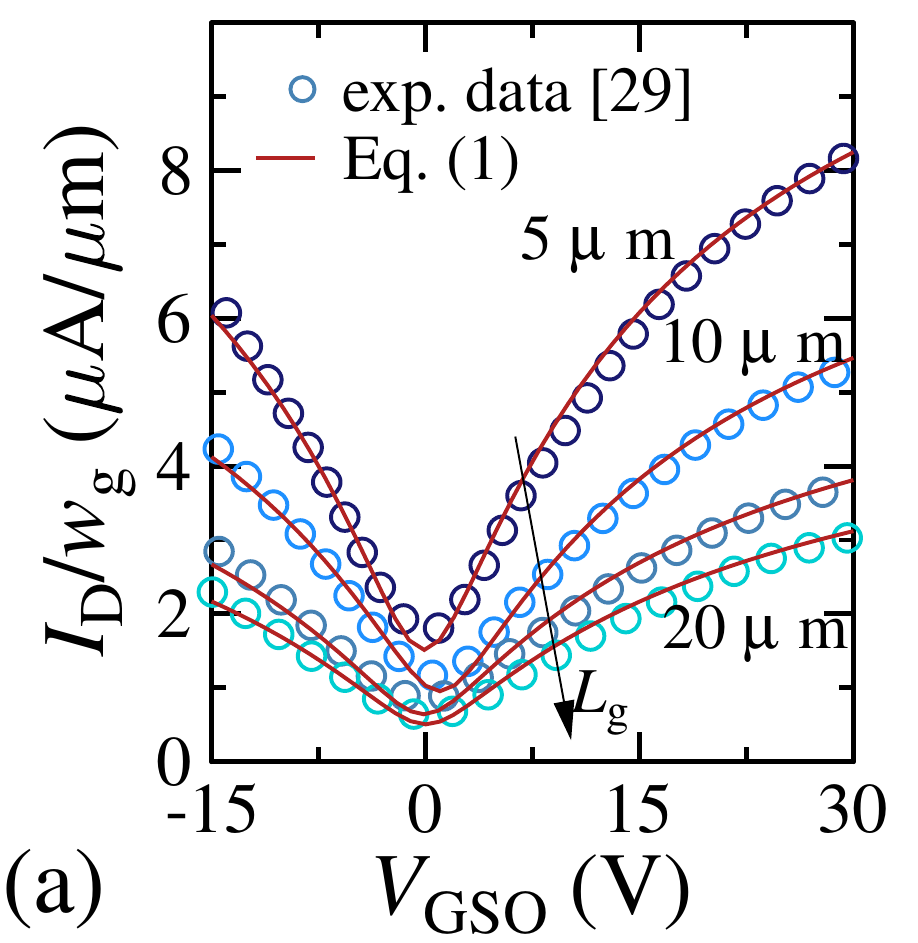}
	\includegraphics[height=0.245\textwidth]{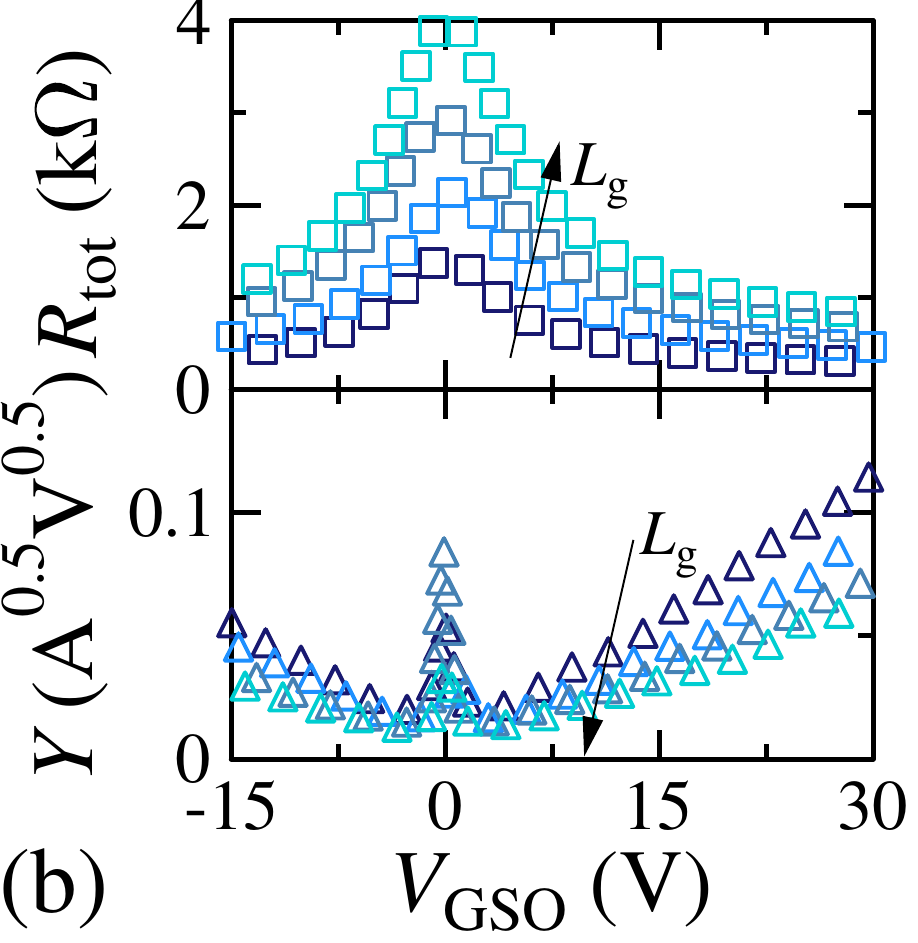} \\
		\includegraphics[height=0.245\textwidth]{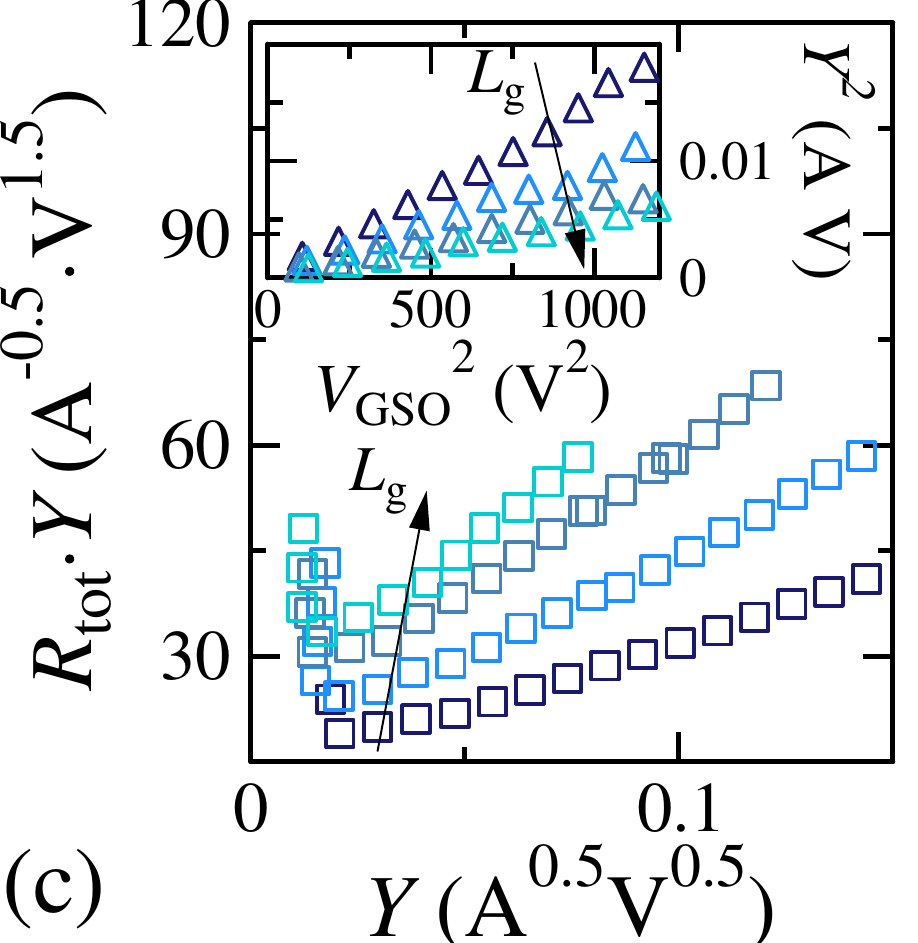}
	\includegraphics[height=0.245\textwidth]{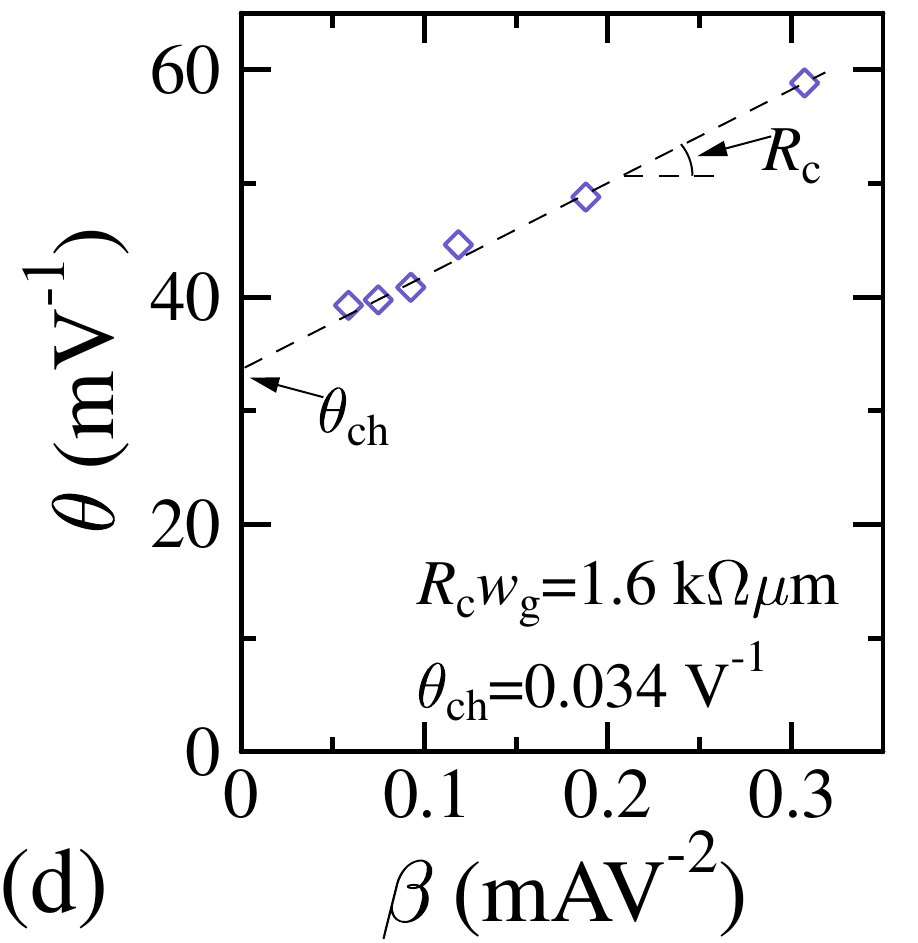} 
	\caption{Extraction methodology presented in this work applied to the GBG GFET technology presented in \cite{GahKat20}. (a) Transfer characteristics, (b) total resistance (top) and $Y$-function (bottom) plots over $V_{\rm GSO}$, (c) $R_{\rm tot}Y$ and $Y^2$ (inset) over $Y$ and $V_{\rm GSO}^2$, respectively and (d) $\theta$ over $\beta$ plot showing the extracted parameters $R_{\rm c}$ and $\theta_{\rm ch}$. Markers represent experimental data and solid lines represent Eq. (\ref{eq:Id}) results by using the extracted parameters. Dashed lines are added as guides for the eyes in order to show the extracted parameters. $V_{\rm DS}=\SI{0.05}{\volt}$ for all plots.}
	\label{fig:IdVg_gfet_B}
\end{figure}

The extraction procedure is depicted in Fig. \ref{fig:IdVg_gfet_B} with experimental data of GBG devices reported elsewhere \cite{GahKat20}. Only the \textit{n}-type region ($V_{\rm GSO}>0$) has been used for the extraction. From the experimental transfer characteristics of GFETs with different $L_{\rm g}$ and at a given $V_{\rm DS}$, plots of $R_{\rm tot}$ and the $Y$-function have been obtained (Fig. \ref{fig:IdVg_gfet_B}(b)). The product of $R_{\rm tot}$ and $Y$ has been ploted over $Y$ (Fig. \ref{fig:IdVg_gfet_B}(c))  in order to obtain the ratio $\theta/\beta$ from the slope of the linear part of each curve (at $Y>Y\vert_{\min(R_{\rm tot}Y)}$), i.e., for each $L_{\rm g}$, as indicated by Eq. (\ref{eq:RtotY}). Similarly, the slopes of the curves in the $Y^2(V_{\rm GSO}^2)$-plot (inset of Fig. \ref{fig:IdVg_gfet_B}(c)) yield the parameter $\beta$. The contact resistance and the intrinsic mobility degradation coefficient are obtained from the $\theta(\beta)$ plot (Fig. \ref{fig:IdVg_gfet_B}(d)). In order to validate the extracted parameters, the model with the extracted parameters succesfully describes the experimental transfer characteristics, provided $V_0$ is obtained, as shown in Fig. \ref{fig:IdVg_gfet_B}(a). 
  
\begin{figure}[!hbt]
	\centering
	\includegraphics[height=0.245\textwidth]{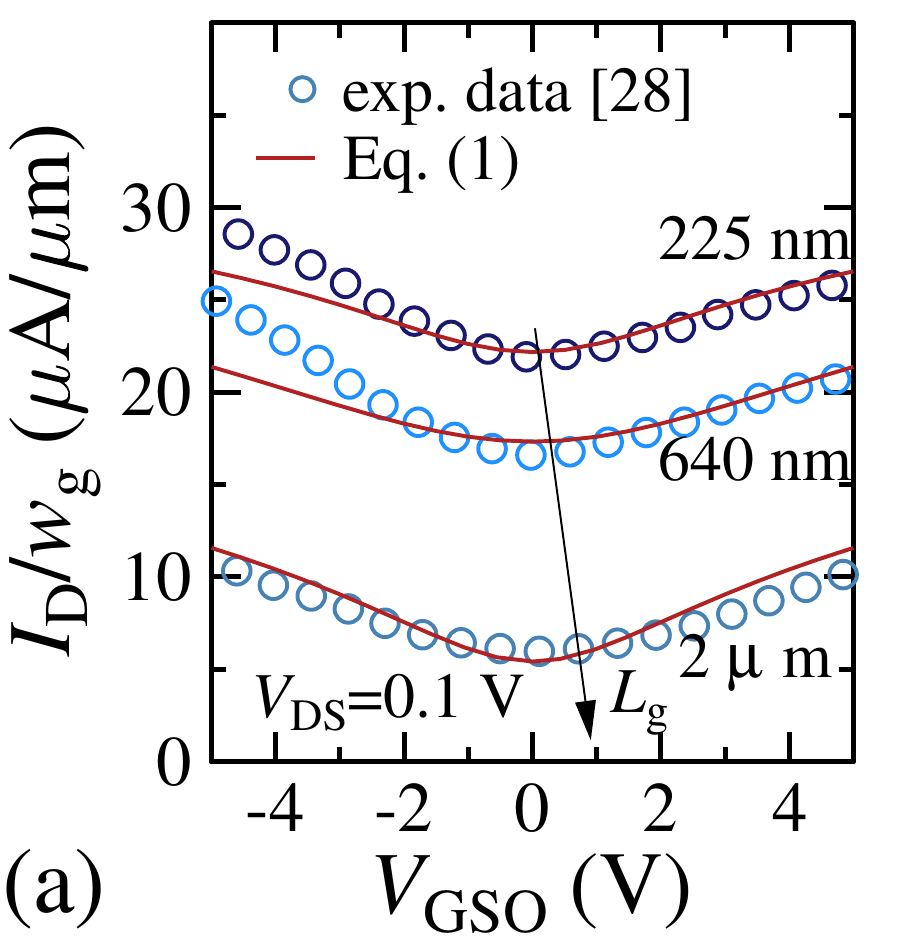}
	\includegraphics[height=0.245\textwidth]{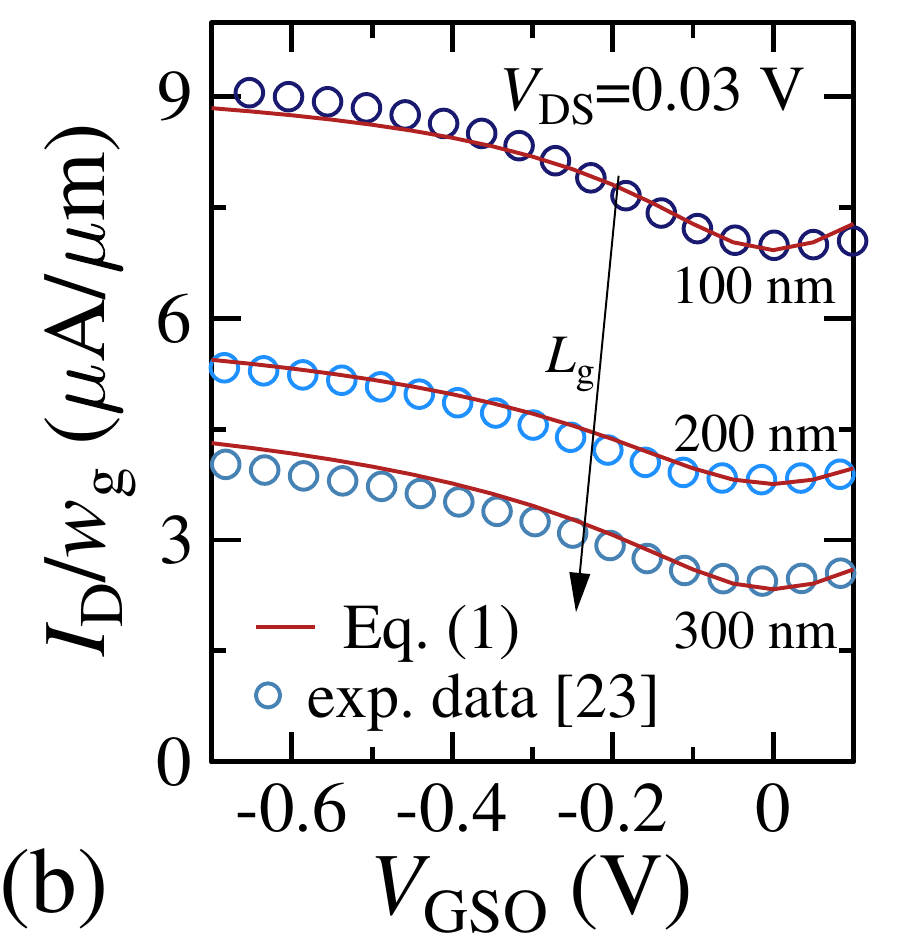} \\
		\includegraphics[height=0.245\textwidth]{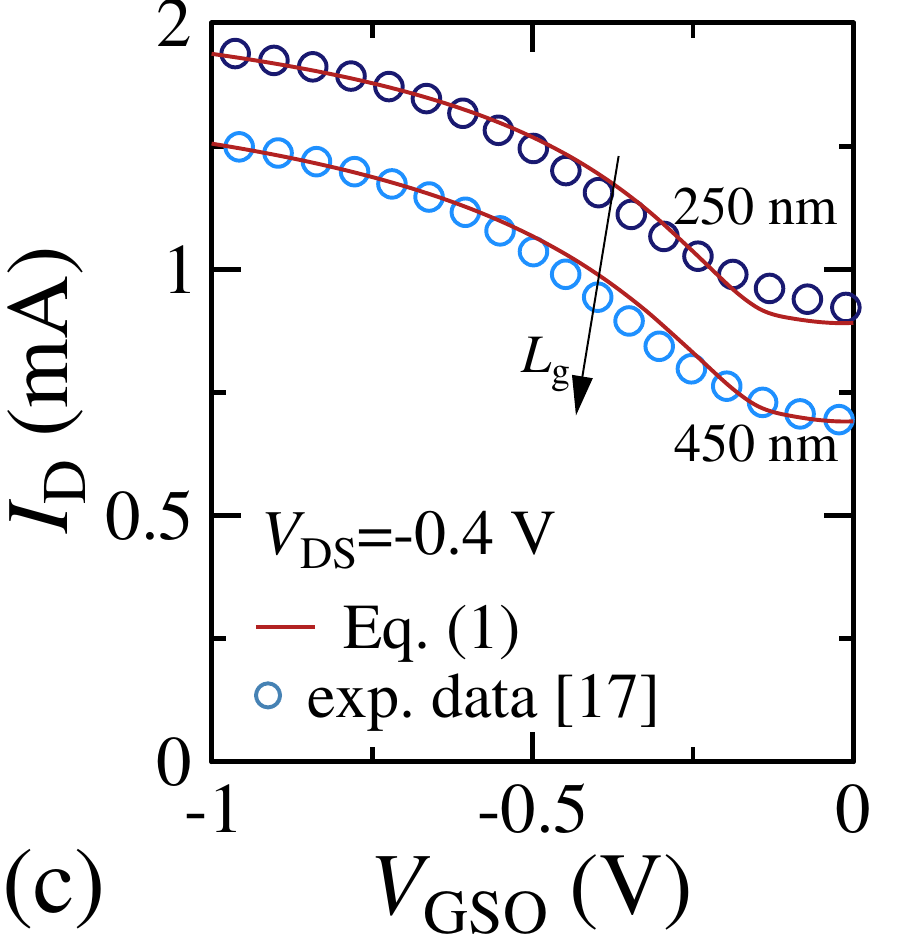}
	\includegraphics[height=0.245\textwidth]{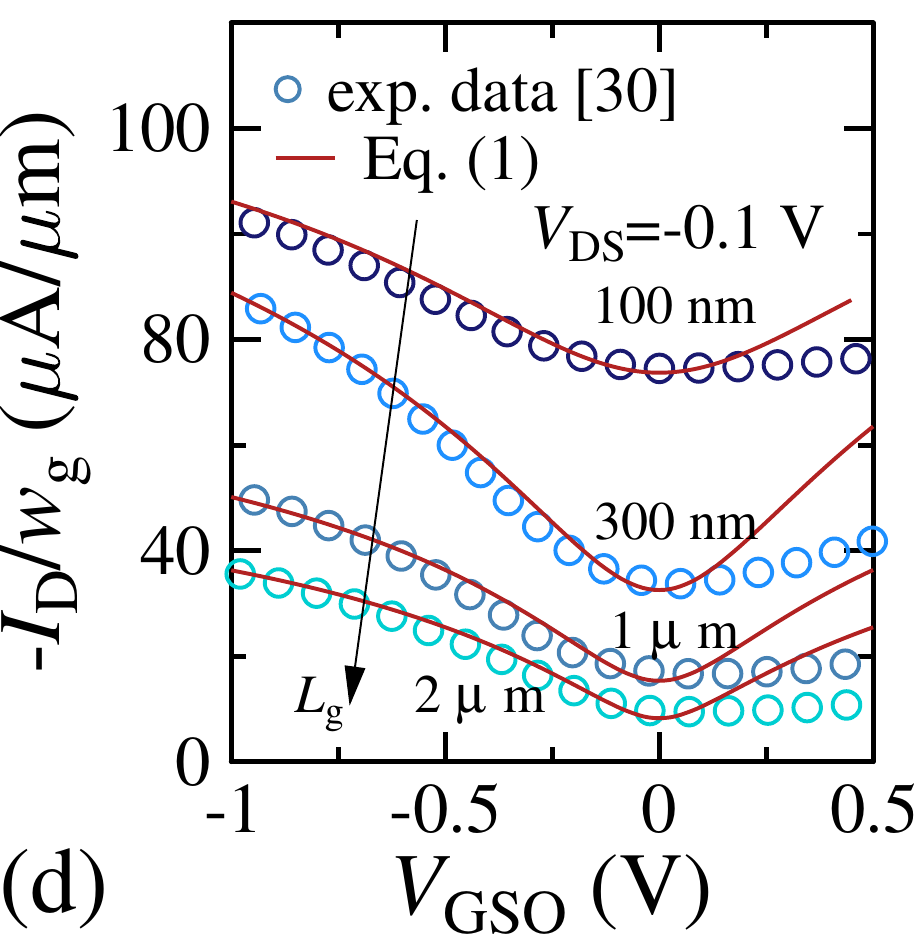} 
	\caption{Transfer characteristics of GFET technologies with different architectures: (a) GBG \cite{HanChe11}, (b) BG \cite{MavWei19}, (c) BG \cite{HanOid13} and (d) TG \cite{BaiLia11}. Markers represent experimental data and solid lines represent modeling results by using the extracted parameters in Eq. (\ref{eq:Id}).}
	\label{fig:IdVg_gfet_all}
\end{figure}

By using the corresponding extracted parameters (cf. Table \ref{tab:results}), the mobility degradation-based transport model describes the experimental transfer characteristics of the other GFET technologies studied here within the bias region where the extraction methodology has been applied in each case as shown in Fig. \ref{fig:IdVg_gfet_all}. The values of $\mu_{\rm app,0}$ for the  shortest devices in the BG technologies presented in this work are of $\SI{2079}{\centi\meter^2\volt^{-1}\second^{-1}}$ for \cite{HanOid13} and of $\SI{2580}{\centi\meter^2\volt^{-1}\second^{-1}}$ for \cite{MavWei19}. These values are similar to the ones reported in the literature: \SIrange{2000}{2500}{} $\SI{}{\centi\meter^2\volt^{-1}\second^{-1}}$ for the former one \cite{HanOid13} and of $\SI{3250}{\centi\meter^2\volt^{-1}\second^{-1}}$, reported with Hall measurements in \cite{WeiZho15}, for the latter \cite{MavWei19}. 

The impact of velocity saturation, out of the scope of this modeling approach, can further improve the model adjustment. E.g., if a simple velocity saturation model for an effective mobility $\mu_{\rm eff}(=\mu / (1+E_{\rm x}/E_{\rm c}))$ is considered, where $\mu$ includes mobility degradation effects as Eq. (1) in \cite{JepAsa21} and $E_{\rm x}$ and $E_{\rm c}$ are the horizontal and critical electric field, respectively, the related velocity saturation parameters can be extracted at high $V_{\rm DS}$ regime provided that all the other parameters, i.e., $\theta$, $R_{\rm c}$ and $\beta$ have been estimated from a lower $V_{\rm DS}$ region with the extraction method presented here. A similar approach for the estimation of velocity saturation effects has been presented elsewhere \cite{MavWei20}. 

The pronounced asymmetry in experimental transfer characteristics (cf. Figs. \ref{fig:IdVg_gfet_all}(a) and (d)) due to different mobility and contact properties observed in the devices of \cite{HanChe11} and \cite{BaiLia11} is not captured by the model. Hence, slight discrepancies are observed in the unipolar region where the extraction has not been performed, i.e., \textit{p}-type in \cite{HanChe11} and \textit{n}-type region in \cite{BaiLia11}. In this case, a new parameter set obtained by applying the extraction method at the bias conditions of interest can be used to overcome the disagreement due to asymmetric device features. For the GBG (TG) device in \cite{HanChe11} (\cite{BaiLia11}), the extraction methodology applied for the \textit{p}-type (\textit{n}-type) region yields values of $R_{\rm c}\cdot w_{\rm g}$ equal to $\SI{0.8}{\kilo\ohm\cdot\micro\meter}$ ($\SI{0.65}{\kilo\ohm\cdot\micro\meter}$) and of $\theta_{\rm ch}$ equal to $\SI{0.04}{\volt^{-1}}$ ($\SI{0.29}{\volt^{-1}}$). The different current levels in each part of the transfer curves (Fig. \ref{fig:IdVg_gfet_all}(a) and (d)) of the device in \cite{HanChe11} (\cite{BaiLia11}) are associated to the lower (higher) $R_{\rm c}\cdot w_{\rm g}$ extracted value for the \textit{p}-type (\textit{n}-type) region in contrast to the one obtained for the \textit{n}-type (\textit{p}-type) region (cf. Table \ref{tab:results}). Additionally, the weaker gate control over the channel observed for the TG device \cite{BaiLia11} at $V_{\rm GSO}>\SI{0}{\volt}$ than at $V_{\rm GSO}<\SI{0}{\volt}$ is quantified by an extracted $\theta_{\rm ch}$ value which is $\sim\SI{3.6}{}$ higher than the one extracted for the \textit{n}-type region.

Regarding the extracted $R_{\rm c}$ values, for the GBG technology with relaxed dimensions, the extracted value of $R_{\rm c}$ (cf. Fig. \ref{fig:IdVg_gfet_B}(b)) is in good agreement with the one obtained by TLM in \cite{GahKat20} considering the bias region where the methods have been applied, i.e., $\SI{9}{\volt} < V_{\rm GSO} < \SI{37}{\volt}$. In the case of the BG GFET reported in \cite{MavWei19}, the slight difference ($\sim\SI{30}{\ohm}$) between the extracted $R_{\rm c}$ value here and the one obtained by the adjustment of a compact model is related to different underlying transport equations. The extraction of $R_{\rm c}$ with this methodology for all kind of device architectures overcomes the challenging fabrication of TLM test structures which are generally limited to GBG devices \cite{Sch06}, specially in graphene technologies \cite{UrbLup20}, \cite{GahKat20}. Values of $\theta_{\rm ch}$ for GFETs have been rarely reported in the literature with other methods \cite{PacJim20}. Regarding the GFETs analyzed here, GBG devices have the lower $\theta_{\rm ch}$ owing to the full-gated channels in contrast to BG and TG devices, and hence, a better overall gate control over the channel is obtained with the former device architecture. It is worth to notice that in the device technologies studied here, $\theta_{\rm ch}$ is not negligible for the definition of $\theta$ as suggested elsewhere for 2D-based devices \cite{ChaZhu14}. The latter is of key importance to determine whether the device performance is dominated by channel or contact phenomena \cite{PacJim20}.

\begin{figure}[!hbt]
	\centering
	\includegraphics[height=0.245\textwidth]{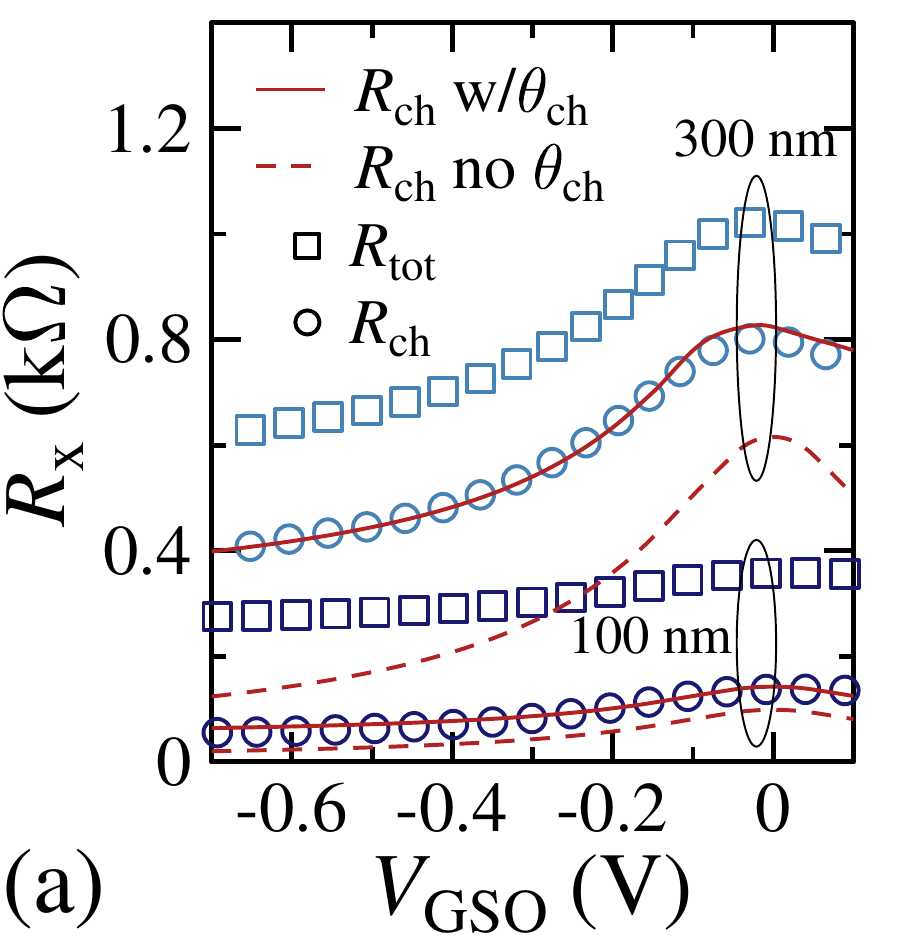}
	\includegraphics[height=0.245\textwidth]{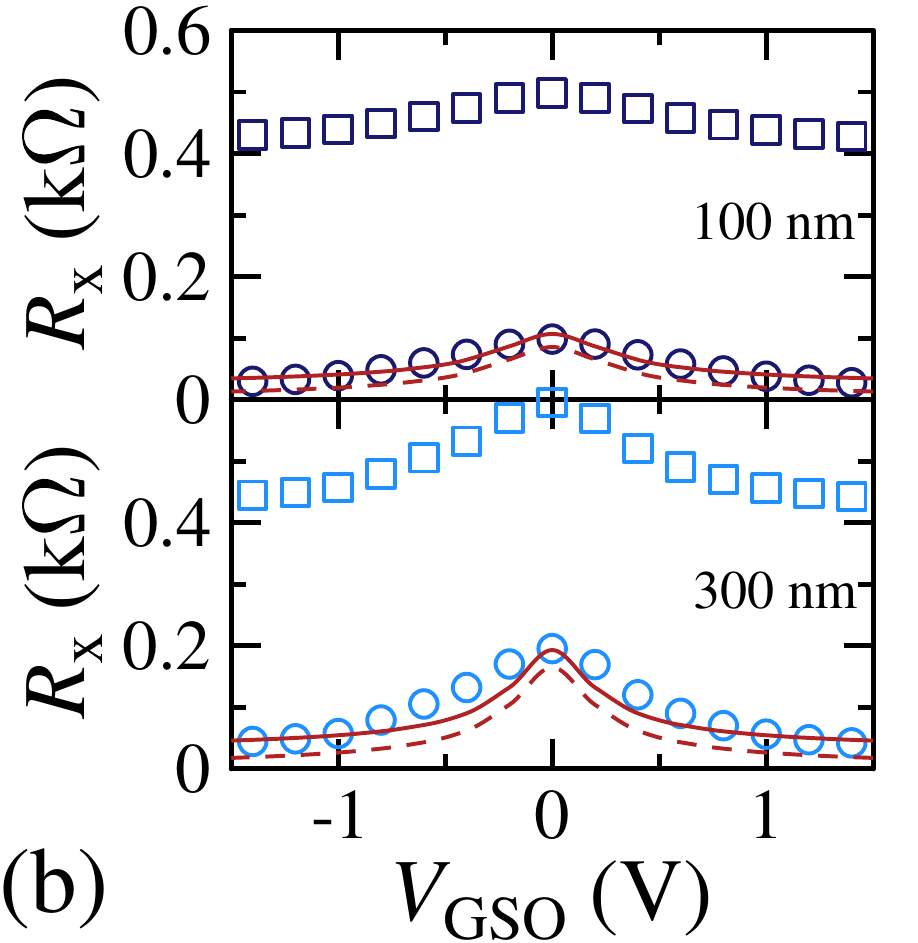} 
	\caption{Resistances of (a) fabricated ($V_{\rm DS}=\SI{0.03}{\volt}$) \cite{MavWei19} and (b) simulated ($V_{\rm DS}=\SI{0.1}{\volt}$) GFETs with different $L_{\rm g}$. Solid (dashed) lines are modeling results of $R_{\rm ch}$ considering (not considering) $\theta_{\rm ch}$. Legend in (a) applies to (b).}
	\label{fig:resist_gfet_all}
\end{figure}

The experimental device channel resistance BG GFET technology reported in \cite{MavWei19}, obtained upon extraction of $R_{\rm c}$, is well described by the modeling approach followed here, i.e., considering $\theta_{\rm ch}$ into its description (c.f. Eq. (\ref{eq:Rtot})), as shown in Fig. \ref{fig:resist_gfet_all}(a). The experimental $R_{\rm tot}$ curves have been included in the plot for comparison purposes. Modeling results of $R_{\rm ch}$ without including $\theta_{\rm ch}$ fail to describe the experimental data: the maximum (minimum) error between model without $\theta_{\rm ch}$ is of $\sim\SI{60}{}\%$ ($\sim\SI{25}{}\%$) in contrast to the $\sim\SI{15}{}\%$ ($\sim\SI{0.5}{}\%$) maximum (minimum) relative error for the $R_{\rm ch}$ model considering $\theta_{\rm ch}$, all values for the region where the parameter extraction has been performed for these devices, i.e., for $V_{\rm GSO}<\SI{0}{\volt}$.

Numerical device simulations (NDS) have been performed in order to further explore the channel resistance description. The simulated GFETs under study have gated-channels (of lengths of \SIlist{100;300}{\nano\meter} and a width of \SI{1}{\micro\meter}) encapsulated in a hexagonal-boron nitride dielectric. Mobility degradation due to vertical fields has been considered. The contact resistance value is set to \SI{400}{\ohm}. The model used in simulations consists in the self-consistent solution of the Poisson's equation together with the current continuity equation, where the properties of carrier transport have been previously extracted from Monte Carlo simulations \cite{PacFei20}, \cite{FeiPas19}. Simulation results, shown in Fig. \ref{fig:resist_gfet_all}(b), reveal that the devices total resistance is mainly due to the injection mechanisms ($R_{\rm c}>R_{\rm ch}$). The extraction method has been applied to the NDS data yielding an $R_{\rm c}$ value of \SI{397}{\ohm} and a $\theta_{\rm ch}$ of \SI{1}{\volt^{-1}}. The channel resistance obtained from the simulation data by $R_{\rm ch}=R_{\rm tot}-\SI{400}{\ohm}$ has been better described by an $R_{\rm ch}$ model considering $\theta_{\rm ch}$ (maximum/minimum error of $\sim\SI{30}{}\%/\SI{3}{}\%$ at the region used for extraction) in contrast to the approaches suggested elsewhere \cite{JepAsa21}, \cite{ChaZhu14} where $\theta_{\rm ch}$ has been neglected as a channel transport parameter (maximum/minimum error of $\sim\SI{55}{}\%/\SI{30}{}\%$ at the region used for extraction). 

Hence, in addition to the conceptual relation of $\theta_{\rm ch}$ to channel phenomena (and corresponding parameters) and the discussion provided above, the results presented here suggest that the term including $\theta_{\rm ch}$ should be accounted for an accurate description of $R_{\rm ch}$. The $L_{\rm g}$ independence of $R_{\rm c}$, implied with the latter approach, has been also confirmed with experimentally calibrated physics-based contact models of GFETs \cite{ChaJim15}, \cite{WanMal21}. In other words, in contrast to a previous proposal where the $R_{\rm c}$ value varies with $L_{\rm g}$, the effect of a gate electrostatics-dependent effective length of the potential step (in the transport direction) at metal-graphene interfaces \cite{ChaJim15}, \cite{XiaPer11} has been associated in this approach to the transport within the channel rather than to contact effects.

\section{Conclusion}

A parameter extraction methodology of a novel model considering mobility degradation effects in GFETs has been provided. The underlying transport model using the extracted parameters, namely $R_{\rm c}$, $\theta_{\rm ch}$ and $\beta$, describes accurately the experimental data of various short- and long-gate GFET technologies at different bias conditions regardless the gate architecture. The latter is of relevance for leveraging the challenging fabrication of test structures for parameter extraction. The engineering approach proposed with this extraction method relies entirely on a mathematical treatment of the experimental transfer characteristic enabled by a mobility degradation-based modeling proposal, i.e., no initial fitting parameters are required. The mobility degradation description enabled by the extracted parameters can be useful for technology evaluation and modeling purposes as shown in this work. A modeling approach including $\theta_{\rm ch}$ in the description of the channel resistance has been proven to yield more accurate results than the case where $\theta_{\rm ch}$ is not considered into $R_{\rm ch}$. The extracted parameters can be included in compact models. The extracted constant value of $R_{\rm c}$ can be exploited in practical small-signal GFET modeling approaches where devices are biased at a certain operation regime.


\begin{thebibliography}{1}

\bibitem{KimNah09} S. Kim, J. Nah, I. Jo, D. Shahrjerdi, L. Colombo, Z. Yao, E. Tutuc, S. K. Banerjee, "Realization of a high mobility dual-gated graphene field-effect transistor with Al$_2$O$_3$ dielectric", \textit{Applied Physics Letters}, vol. 94, no. 9, Feb. 2009. DOI: 10.1063/1.3077021


\bibitem{LiaBai10} L. Liao, J. Bai, Y. Qu, Y. Lin, Y. Li, Y. Huang, X. Duan, "High-k oxide nanoribbons as gate dielectrics for high mobility top-gated graphene transistors", \textit{Proceedings of the National Academy of Sciences of the United States of America}, vol. 107, no. 15, pp. 6711-6715, Apr. 2010. DOI: 10.1073/pnas.0914117107 





\bibitem{WuTse15} Y.-H. Wu, P.-Y. Tseng, P.-Y. Hsieh, H.-T. Chou, N.-H. Tai, "High Mobility of Graphene-Based Flexible Transparent Field Effect Transistors Doped with TiO2 and Nitrogen-Doped TiO2", \textit{ACS Applied Materials \& Interfaces}, vol. 7, no. 18, pp. 9453-9461, Apr. 2015. DOI: 10.1021/am508996r


\bibitem{MavWei20} N. Mavredakis, W. Wei, E. Pallecchi, D. Vignaud, H. Happy, R. G. Cortadella, N. Schaefer, A. B. Calia, J. A. Garrido, D. Jimenez, "Low-Frequency Noise Parameter Extraction Method for Single-Layer Graphene FETs", \textit{IEEE Transactions on Electron Devices}, vol. 67, no. 5, pp. 2093-2099, May 2020. DOI: 10.1109/TED.2020.2978215


\bibitem{JepAsa21} K. Jeppson, M. Asad, J. Stake, "Mobility Degradation and Series Resistance in Graphene Field-Effect Transistors", \textit{IEEE Transactions on Electron Devices}, vol. 68, no. 6, pp. 3091-3095, Jun. 2021. DOI: 10.1109/TED.2021.3074479

\bibitem{ZhuPer09} W. Zhu, V. Perebeinos, M. Freitag, and P. Avouris, “Carrier scattering, mobilities, and electrostatic potential in monolayer, bilayer, and trilayer graphene”,  \textit{Physical Review B}, vol. 80, no. 23, 235402, Dec. 2009. DOI: 10.1103/PhysRevB.80.235402.

\bibitem{DorBae10} V. E. Dorgan, M.-H. Bae, E. Pop, “Mobility and saturation velocity in graphene on SiO2”, \textit{Applied Physics Letters}, vol. 97, no. 8, Art. no. 082112, Aug. 2010. DOI: 10.1063/1.3483130.


\bibitem{ZhoZha15} H. Zhong, Z. Zhang, H. Xu, C. Qiu, and L.-M. Peng, “Comparison of mobility extraction methods based on field-effect measurements for graphene”, \textit{AIP Advances}, vol. 5, no. 5, Art. no. 057136, May 2015. DOI: 10.1063/1.4921400.

\bibitem{MerBor72} G. Merckel, J. Borel, N. Z. Cupcea, "An accurate large-signal MOS transistor model for use in computer-aided design", \textit{IEEE Transactions on Electron Devices}, vol. 19, no. 5, pp. 681–690, 1972. DOI: 10.1109/T-ED.1972.17474

\bibitem{Ris83} L. Risch, "Electron mobility in short-channel MOSFET's with series resistances", \textit{IEEE Transactions on Electron Devices}, vol. 30, no. 8, pp. 959-961, Aug. 1983. DOI: 10.1109/T-ED.1983.21246.

\bibitem{EnzVit06} C. C. Enz, E. A. Vittoz, "Charge-based MOS Transistor Modeling", John Wiley \& Sons, 2006.


\bibitem{UrbLup20} F. Urban, G. Lupina, A. Grillo, N. Martucciello, A. Di Bartolomeo, "Contact resistance and mobility in back-gate graphene transistors", \textit{Nano Express}, vol. 1, no. 1, 010001, Mar. 2020. DOI: 10.1088/2632-959X/ab7055

\bibitem{PacFei20} A. Pacheco-Sanchez, P. C. Feijoo, D. Jiménez, “Contact resistance extraction of graphene FET technologies based on individual device characterization”, \textit{Solid-State Electronics}, vol. 172, 107882, Oct., 2020. DOI: 10.1016/j.sse.2020.107882

\bibitem{PacJim20} A. Pacheco-Sanchez, D. Jiménez, "Accuracy of Y-function methods for parameters extraction of two-dimensional FETs across different technologies", \emph{Electronics Letters}, vol. 56, no. 18, pp. 942-945, Sep. 2020. DOI: 10.1049/el.2020.1502

\bibitem{PacJim21} A. Pacheco-Sanchez, D. Jiménez, "A contact resistance extraction method of 2D-FET technologies without test structures", in Proc. \textit{IEEE Spanish Conference on Electron Devices (CDE)}, pp. 19-22, 2021. DOI: 10.1109/CDE52135.2021.9455755

\bibitem{Sch06} D. Schroder, "Semiconductor material and device characterization", Wiley-IEEE Press, 2006.

\bibitem{HanOid13} S.-J. Han, S. Oida, K. A. Jenkins, D. Lu,  Y. Zhu, "Multifinger Embedded T-Shaped Gate Graphene RF Transistors With High fMAX/fT Ratio", \textit{IEEE Electron Device Letters},vol. 34, no. 10, pp. 1340-1342, Oct. 2013. DOI: 10.1109/LED.2013.2276038

\bibitem{LeeKim19} S. K. Lee, Y. J. Kim, S. Heo, W. Park, T. J. Yoo, C. Cho, H. J. Hwang, B. H. Lee, "Advantages of a buried-gate structure for graphene field-effect transistor", \textit{Semiconductor Science and Technology}, vol. 34, 055010, Apr. 2019. DOI: 10.1088/2053-1583/aba449

\bibitem{ChaJim15} F. A. Chaves, D. Jiménez, A. A. Sagade, W. Kim, J. Riikonen, H. Lipsanen, D. Neumaier, "A physics-based model of gate-tunable metal–graphene contact resistance benchmarked against experimental data", \textit{2D Materials}, vol. 2, no. 2, 025006, May 2015. DOI: 10.1088/2053-1583/2/2/025006

\bibitem{WanMal21} B. Wang , M. W. Malik, Y. Yan, V. Kilchytska, Y. Zeng, D. Flandre, J.-P. Raskin, "A Physical Model of Contact Resistance in Ti-Contacted Graphene-Based Field Effect Transistors", \textit{IEEE Transactions on Electron Devices}, vol. 68, no. 2, pp. 892-898, Feb. 2021. DOI: 10.1109/TED.2020.3046166

\bibitem{HaoCab85} C. Hao, B. Cabon-Till, S. Cristoloveanu, G. Ghibaudo, "Experimental determination of short-channel MOSFET parameters", \emph{Solid-State Electronics}, vol. 28, no. 10, pp. 1025-1030, 1985. DOI: 10.1016/0038-1101(85)90034-6

\bibitem{Ghi88} G. Ghibaudo, "New method for the extraction of MOSFET parameters", \textit{Electronics Letters}, vol. 24, no. 9, pp. 543-545, 1988. DOI: 10.1049/el:19880369 


\bibitem{MavWei19} N. Mavredakis, W. Wei, E. Pallecchi, D. Vignaud, H. Happy, R. G. Cortadella, A. B. Calia, J. A. Garrido, D. Jiménez, "Velocity Saturation Effect on Low Frequency Noise in Short Channel Single Layer Graphene Field Effect Transistors", \textit{ACS Applied Electronic Materials}, vol. 1, pp. 2626-2636, 2019. DOI: 10.1021/acsaelm.9b00604 

\bibitem{Jai88} S. Jain, "Measurement of threshold voltage and channel length of submicron MOSFETs", \emph{IEE Proceedings I (Solid-State and Electron Devices)}, vol. 135, no. 6, pp. 162–164, 1988. DOI: 10.1049/ip-i-1.1988.0029



\bibitem{PacRam21} A. Pacheco-Sanchez, J. N. Ramos-Silva, E. Ramírez-García, D. Jiménez “A Small-Signal GFET Equivalent Circuit Considering an Explicit Contribution of Contact Resistances”, \textit{IEEE Microwave and Wireless Components Letters}, vol. 31, no. 1, pp. 29-32, Jan. 2021.
DOI: 10.1109/LMWC.2020.3036845

\bibitem{HamAsa21} A. Hamed, M. Asad, M. -D. Wei, A. Vorobiev, J. Stake, R. Negra, "Integrated 10-GHz Graphene FET Amplifier", \textit{IEEE Journal of Microwaves}, Jun. 2021. DOI: 10.1109/JMW.2021.3089356.

\bibitem{TreTre17} R. Trevisoli, R. Trevisoli Doria, M. de Souza, S. Barraud, M. Vinet, M. Cassé, G. Reimbold, O. Faynot, G. Ghibaudo, M. A. Pavanello, "A New Method for Series Resistance Extraction of Nanometer MOSFETs", \textit{IEEE Transactions on Electron Devices}, vol. 64, no. 7, pp. 2797-2803, Jul. 2017. DOI: 10.1109/TED.2017.2704928


\bibitem{HanChe11} S.-J. Han, Z. Chen, A. A. Bol, Y. Sun, "Channel-Length-Dependent Transport Behaviors of Graphene Field-Effect Transistors", \textit{IEEE Electron Device Letters}, vol. 32, no. 6, pp. 812-814, Jun. 2011. DOI: 10.1109/LED.2011.2131113

\bibitem{GahKat20} A. Gahoi, S. Kataria, F. Driussi, S. Venica, H. Pandey, D. Esseni, L. Selmi, M. C. Lemme, ”Dependable Contact Related Parameter Extraction in Graphene-Metal Junctions”, \textit{Advanced Electronic Materials}, vol. 6, 2000386, Sep. 2020. DOI: 10.1002/aelm.202000386

\bibitem{BaiLia11} J. Bai, L. Liao, H. Zhou, R. Cheng, L. Liu, Y. Huang, X. Duan, "Top-Gated Chemical Vapor Deposition Grown Graphene Transistors with Current Saturation", \textit{Nano Letters}, vol. 11, pp. 2555-2559, May 2011. DOI:  10.1021/nl201331x



\bibitem{ChaZhu14} H.-Y. Chang, W. Zhu, D. Akinwande, ”On the mobility and contact resistance evaluation for transistors based on MoS2 or two dimensional semiconducting atomic crystals”, \textit{Applied Physics Letters}, vol. 104, p. 113504, 2014. DOI: 10.1063/1.4868536

\bibitem{WeiZho15} W. Wei, X. Zhou, G. Deokar, H. Kim, M. M. Belhaj, E. Galopin, E. Pallecchi, D. Vignaud, H. Happy, "Graphene FETs With Aluminum Bottom-Gate Electrodes and Its Natural Oxide as Dielectrics", \textit{IEEE Transactions on Electron Devices}, vol. 62, no. 9, pp. 2769-2773, Aug. 2015. DOI: 10.1109/TED.2015.2459657

\bibitem{FeiPas19} P. C. Feijoo, F. Pasadas, J. M. Iglesias, E. M. Hamham, R. Rengel, D. Jiménez, "Radio Frequency Performance Projection and Stability Tradeoff of h-BN Encapsulated Graphene Field-Effect Transistors", \textit{IEEE Transactions on Electron Devices}, vol. 66, no. 3, pp. 1567-1573, Mar. 2019. DOI: 10.1109/TED.2018.2890192

\bibitem{XiaPer11} F. Xia, V. Perebeinos, Ym. Lin, Y. Wu, P. Avouris, "The origins and limits of metal–graphene junction resistance", \textit{Nature Nanotechnology}, vol. 6, pp. 179–184, Feb. 2011. DOI: 10.1038/nnano.2011.6





\end{thebibliography}
\end{document}